\documentclass[12pt]{article}
\pdfoutput=1
\usepackage{amsmath,amsfonts,amssymb}
\usepackage{graphicx}

\newcommand{\be}{\begin{equation}}
\newcommand{\ee}{\end{equation}}
\newcommand{\ba}{\begin{eqnarray}}
\newcommand{\ea}{\end{eqnarray}}

\newcommand{\SU}{\mathrm{SU}}

\newtheorem{theorem}{Theorem}[section]
\newtheorem{lemma}[theorem]{Lemma}

\def\bR {\mathbb{R}}

\def\ep {\epsilon}

\def\be{\begin{equation}}
\def\ee{\end{equation}}
\def\bea{\begin{eqnarray}}
\def\eea{\end{eqnarray}}

\newcommand{\tr}{{\rm tr\,}}

\newcommand{\nn}{\nonumber}

\newcommand{\la}{\langle}
\newcommand{\ra}{\rangle}
\newcommand{\nc}{\newcommand}
\nc{\cH}{\mathcal{H}}

\usepackage{color}

\begin{document}
\newcommand{\todo}[1]{{\em \small {#1}}\marginpar{$\Longleftarrow$}}   
\newcommand{\labell}[1]{\label{#1}\qquad_{#1}} 


\begin{titlepage}

\begin{center}
\centerline{{\Large{\bf{From state distinguishability to effective bulk locality}}}}
\vspace{4mm}

{\large{{\bf Nima Lashkari\footnote{e-mail: nimal@stanford.edu }$^{,a,b}$ and Joan Sim\'on\footnote{e-mail:
j.simon@ed.ac.uk}$^{,c}$ }}}
\\

\vspace{5mm}

\bigskip\medskip
\begin{center}
{$^a$ \it Stanford Institute for Theoretical Physics and Department of Physics,\\
Stanford University, Stanford, CA 94305-4060, USA}\\
\smallskip
{$^b$ \it Department of Physics and Astronomy, University of British Columbia,\\
6224 Agricultural Road,
Vancouver, B.C., V6T 1W9, Canada } \\
\smallskip
{$^c$ \it School of Mathematics and Maxwell Institute for Mathematical Sciences,\\
King's Buildings, Edinburgh EH9 3JZ, United Kingdom}\\
\smallskip
\end{center}
\vfil

\end{center}
\setcounter{footnote}{0}

\begin{abstract}
\noindent
We provide quantitative evidence that the emergence of an effective notion of spacetime locality in black hole physics is due to restricting to the subset of observables that are unable to resolve black hole microstates from the maximally entangled state. We identify the subset of observables in the full quantum theory that can distinguish microstates, and argue that any measurement of such observables involves either long times or large energies, both signaling the breaking down of effective field theory where locality is manifest. We discuss some of the implications of our results for black hole complementarity and the existence of black hole interiors.

\end{abstract}

\end{titlepage}

\newpage

\section{Introduction}

One of the main open questions in black hole physics is the compatibility of unitarity in the entire quantum theory with the emergence of some notion of spacetime locality, the latter being manifest in any effective field theory (EFT) description.

In this paper we explore the emergence of some notion of {\it effective bulk locality} in light of the connection between the resolution of the information paradox, and the modern formulation of  thermalization in terms of {\it entanglement} and {\it typicality}. In a nutshell, just as the emergence of {\it effective thermalization} in quantum statistical mechanics requires some notion of coarse-graining, we propose that the emergence of effective bulk locality is due to the restriction to a subset of observables which cannot resolve black hole microstates from their maximally entangled averages.

Using tools from quantum information theory, we prove that a given observable ${\cal O}$, or a collection of them, cannot distinguish a random pure microstate in a microcanonical ensemble ${\cal H}_E$ of dimension $d_E$ from the maximally entangled state $\Omega_E=\frac{\mathbb{I}_E}{d_E}$ unless the number of different outcomes of the operator $N({\cal O})$ scales as $\sqrt{d_E}$. Furthermore, whenever $N({\cal O})\sim\sqrt{d_E}$, we prove, in a simple quantum mechanical model, that any quantum measurement would require either a very long time or involve a very large energy to achieve the accuracy required to distinguish these states. Either way, this points to a breakdown in the EFT description. Alternatively, we show that any measurement involving a finite amount of resources is necessarily coarse-grained.

In an AdS/CFT set-up, our results provide mathematical evidence that the operators belonging to the bulk low energy effective theory are coarse-grained, in the sense that they are unable to distinguish random pure states from the maximally mixed one unless you wait for exponentially long time. Furthermore, they also support the idea that only low point correlators of such operators admit a {\it local} bulk semi-classical interpretation compatible with the EFT description. Taking this perspective,  we further comment on some implications for quantum gravity: mainly the consistency with black hole complementarity and the relation with recent discussions regarding the (non-)existence of a classical black hole interior.

\section{Black holes vs Quantum Statistical Mechanics}

Hawking established that black holes radiate with a thermal spectrum \cite{hawking} by performing a 2-point function calculation in a framework that we shall henceforth refer to as effective field theory (EFT): he considered a bulk quantum field $\phi(x)$ propagating in a non-dynamical black hole background
\begin{equation}
  \langle \psi_{\text{BH}} |\phi (x_1)\,\phi(x_2) | \psi_{\text{BH}}\rangle_{\text{EFT}}\,,
\end{equation}
where the subscript EFT refers to the replacement of the black hole quantum state $| \psi_{\text{BH}}\rangle$ by the classical black hole background.

This description is manifestly local, a property that is expected to hold due to the equivalence principle, but is hard to derive in any quantum theory of gravity. Locality appears to be in contradiction with the unitarity of quantum mechanics, and this is commonly referred to as the {\it information paradox}. One aspect of this paradox is directly related to the {\it problem of thermalization} in quantum mechanics: given a {\it closed} quantum system in an initial pure state, unitary evolution never gives rise to a mixed state such as the Gibbs state. The latter is a fundamental problem for the foundations of quantum statistical mechanics (QSM).

How do we understand thermalization then? In QSM, thermalization is always viewed as a consequence of {\it coarse-graining} the description of our physical system. In modern language, one expects  {\it entanglement} and {\it typicality} to explain why QSM is an accurate description of nature at a macroscopic level. For example, consider a composite Hilbert space ${\cal H}={\cal H}_S\otimes {\cal H}_B$, where ${\cal H}_S$ is the subspace that we will measure and ${\cal H}_B$ stands for the bath. Given an initial quantum state $\rho\in {\cal H}$, its reduced density matrix $\rho_S$
\begin{equation}
  \rho_S = \text{tr}_{{\cal H}_B} \rho
\end{equation}
is the only information that we have access to. Quantum entanglement is responsible for encoding the information about correlations between the subsystems whereas typicality is responsible for the apparent universality in our measurements. More precisely, if the dimensions of these Hilbert spaces satisfy $d_S\ll \sqrt{d_B}$, in the thermodynamic limit $d_B\to \infty$, the deviation  of  $\rho_S$ from the maximally entangled state $\Omega_S = \frac{\mathbb{I}_S}{d_S}\in {\cal H}_S$ is suppressed \cite{popescu-1}. That is
\begin{equation}
  \| \rho_S - \Omega_S \| \to 0 \quad \text{when} \quad \frac{d_S}{\sqrt{d_B}} \to 0
\end{equation}
where $\|\hdots \|$ stands for trace distance.\footnote{$\|A\|=\text{tr}(|A|)=\sum_i|a_i|$ where $a_i$'s are eigenvalues of $A$.}

In this work we take the perspective that such coarse-graining in holographic field theories is responsible for the emergence of a notion of {\it effective locality} (and consequently, the causal structure of spacetime), which is an assumption in Hawking's calculation. To advance our main philosophy, bulk locality emerges only for the subset of observables that fail to distinguish a thermal state, or $\Omega_S$ in a microcanonical set-up, from the actual microstate of the black hole. In gravitational physics having holographic duals, the coarse-graining appears as a result of restricting to the subset of low energy observables ${\mathcal A}_{\text{low}}$ belonging to the effective field theory. 
These correspond to the light sector of the holographic theory in the terminology introduced in \cite{joe}. We will discuss this point more thoroughly in section \ref{sec:lessons}.

With this motivation in mind, the main task we undertake is to identify the subset of observables ${\cal O}$ that can distinguish pure microstates $\rho_i=|\psi_i\rangle\langle \psi_i|$ from the thermal density matrix $\rho_{\text{BH}}$ (the semi-classical description of the black hole)
\begin{equation}
  \text{tr}\left(\rho_i\,{\cal O}\right) \neq \text{tr}\left(\rho_{\text{BH}}\,{\cal O}\right)\,,
\end{equation}
in a given large but {\it finite} dimensional Hilbert space $\mathcal{H}$.\footnote{Strictly speaking, our proofs involve the maximally entangled state rather than the Gibbs state. Extending our tools for the canonical ensemble is a hard open problem we will not attempt here.} In an exact quantum theory of gravity, such as in an AdS/CFT context, we quantify the deviations of correlators computed in black hole microstates $|\psi_i\rangle$ from the effective field theory answer. These deviations are expected to be of the form \cite{nima,review}
\begin{equation}
  \langle \psi_i |{\cal O}_1(x_1) \dots {\cal O}_N(x_N)|\psi_i \rangle_{\text{exact}} =  \langle \psi_{\text{BH}} |{\cal O}_1(x_1) \dots {\cal O}_N(x_N)|\psi_{\text{BH}} \rangle_{\text{exact}} + {\cal O}\left(e^{-(S_\text{BH}-N)}\right)
\label{eq:effect}
\end{equation}
where $e^{S_\text{BH}}=d_E$ is the dimension of the microcanonical Hilbert space ${\cal H}_E$ of black hole microstates. The $e^{-S_\text{BH}}$ exponential suppression is expected from semiclassical gravity \cite{juan} and statistical considerations \cite{mukund}. The last term in the right hand side is currently a vague way of parameterising the fact that depending on the nature of the correlator, i.e. the number $N$ of {\it insertions} and the properties of the individual {\it probe} operators ${\cal O}_i$, such corrections may not be subleading \cite{babel,mukund,review}. Our main goal is to make this statement precise in quantum mechanics. 

Even though our analysis is in the full quantum theory, it concerns the robustness of Hawking's EFT framework because whenever thermal answers are not sharply peaked due to large variances, the notion of bulk locality is not expected to hold.  We stress this breaking down in the EFT description is, a priori, on top of the well established fact that low point correlators, such as the 2-pt function in Hawking's original calculation, must break down at large times, i.e. for times of the order of the black hole evaporation time scale or the Poincar\'e recurrence time. The latter can be made particularly precise in an AdS/CFT context \cite{juan}.

\section{Distinguishing quantum states}
\label{sec:3}

Consider a finite dimensional subspace ${\cal H}_E \subset {\cal H}$ of dimension $d_E$ consisting of all pure states $\psi=|\psi \rangle\la \psi|$ that live in the microcanonical ensemble of energy $[E-\delta E,\,E+\delta E]$. We will assume the Hamiltonian describing the unitary time evolution of the system has {\it non-degenerate energy gaps}. This is a condition on the spectrum stating that the equality $E_k - E_l = E_m - E_n$ can only be satisfied either by $E_k=E_l$ and $E_m=E_n$, or by $E_k=E_m$ and $E_l=E_n$.\footnote{This assumption is believed to be correct for almost all interacting Hamiltonians. The addition of an arbitrarily small random term to the Hamiltonians lifts all degeneracies. All our results continue to hold as long as no energy gap is highly degenerate \cite{short-2}.}

In this section, we identify the necessary condition for a set of observables ${\cal A}$ to distinguish a random pure state $\psi \in {\cal H}_E$ from the maximally mixed state in ${\cal H}_E$. We argue that an actual quantum mechanical measurement of any such observable  requires large resources, i.e. very long times or very high energies. Alternatively, any finite time and energy resourced quantum measurement is equivalent to a {\it coarse-grained} observable for which random pure states still appear entangled.

\subsection{Expectation value of operators}
\label{sec:3.1}

One possibility to quantify the difference between quantum states $\psi \in{\cal H}_E$ is to measure the expectation value of some operator $A$. We can study this either at fixed time, by averaging over the entire set of states, or by averaging over time.

To study these questions, it is instructive to think of $\la\psi|A|\psi\ra$ as a random variable $X$ with uniform distribution on $\cH_E$ or over the positive real line $\bR^+$, respectively \cite{reimann}. \footnote{For a proper definition of the different ensemble averages discussed below, see appendix \ref{app:a}.} Applying Chebyshev's inequality
\begin{equation}
  \text{Prob}\left(|X - \langle X\rangle| \geq \delta x\right) \leq \frac{\sigma_X^2}{\delta x^2}
\label{eq:cheb}
\end{equation}
provides an upper bound on the probability for $X$ to differ from its expectation value $\langle X \rangle$ by some quantity $\delta x$.\footnote{The bound \eqref{eq:cheb} is only meaningful whenever $\frac{\sigma_X^2}{\delta x^2}\leq 1$.}

\paragraph{Ensemble average:} Consider a random state $\psi \in {\cal H}_E$. Since the random variable $X_\psi=\la\psi|A|\psi\ra$ is uniformly distributed over ${\cal H}_E$, its average value $\la A \ra_E$ will equal
\begin{equation}
  \la A \ra_E = \frac{1}{d_E}\text{tr} A = \text{tr}\left(A\Omega_E\right) \quad \text{with} \quad \Omega_E = \frac{\mathbb{I}_E}{d_E}\,.
\label{eq:ensav}
\end{equation}
Thus, the ensemble average always equals the expectation value of the operator $A$ in the maximally entangled state $\Omega_E$ in $\cH_E$.

The variance of the random variable  \cite{lloyd,mukund}
\begin{equation}
  \sigma^2_{A}\equiv \left\langle \left(\langle\psi |A|\psi\rangle -  \langle A\rangle_\psi\right)^2 \right\rangle_\psi = \frac{1}{d_E+1}\left(\frac{1}{d_E}\text{tr}(A^{2}) - \left(\frac{\text{tr}(A)}{d_E}\right)^2\right)
\label{eq:rnm}
\end{equation}
is inversely proportional to the ensemble dimension, i.e. exponentially suppressed in the entropy of the system since, $(d_E+1)^{-1}\approx e^{-S_E}$. 

Chebyshev's inequality \eqref{eq:cheb} provides an upper bound on the probability of observing a large deviation from the averaged expectation values
\begin{equation}
\label{corr}
  \text{Prob}\left(|\la\psi|A|\psi\ra-\tr(A\:\Omega_E)|\geq \delta a\right)\leq \frac{\sigma^2_A}{(\delta a)^2}\,.
\end{equation}

\paragraph{Time average:} Given the existence of recurrences in finite dimensional Hilbert spaces, time evolution of any typical state $\psi$ becomes arbitrarily close to any other state. It is important to determine the probability that a given state differs from its equilibrium configuration at a given instant of time. To study this, we choose a different ensemble and define a random variable $X^\psi_t=\la\psi(t)|A|\psi(t)\ra$ with uniform distribution over $t\in[0,\infty)$. Given an initial random state expanded in energy eigenbasis
\begin{equation}
  |\psi(0)\rangle = \sum_E c_E |E\rangle \in\cH_E\,, \quad \text{with} \quad \sum_E |c_E|^2 = 1\,,
\label{eq:sphere}
\end{equation}
it evolves to $|\psi(t) \rangle = \sum_E e^{-iE t}\,c_E |E\rangle$. Its time-average equals
\begin{equation}
  \omega_\psi\equiv \langle \psi(t)\rangle_t = \lim_{T\to\infty}\frac{1}{T} \int_0^Tdt\:|\psi(t)\ra\la\psi(t)| = \sum_E |c_E|^2 |E\rangle\langle E|\,.
\end{equation}
Notice that in this ensemble the {\it equilibrium} configuration $\omega_\psi$ depends explicitly on the initial state $|\psi(0)\ra$ and is obtained by setting to zero all off-diagonal elements in $|\psi(t)\ra\la\psi(t)|$ in the energy eigenbasis. The latter is due to the dephasing occurring when averaging over time.

The variance $\sigma_A^2(\psi)$ in this random variable is
\begin{eqnarray}
\label{sigma}
  \sigma_A^2(\psi)&=&\la \left(\tr(A\:\psi(t))-\tr(A\:\omega_\psi)\right)^2\ra_t \nn \\
  &=& \sum_{E,E'}|c_E|^2|c_{E'}|^2\:|\la E'|A|E\ra|^2-\sum_E |c_E|^4\:|\la E|A|E\ra|^2\,,
\end{eqnarray}
where we assumed the hamiltonian has a non-degenerate energy gap spectrum. Chebyshev's inequality implies
\begin{equation}
\label{chebysh_time}
  \text{Prob}\left(|\la\psi(t)|A|\psi(t)\ra-\tr(A\:\omega_\psi)|\geq \delta a\right)\leq \frac{\sigma^2_A(\psi)}{(\delta a)^2}\,.
\end{equation}  

\paragraph{Ensemble vs time averages:} The equilibrium configuration $\omega_\psi$ depends on the initial state and one can equally ask how large the probability is for a random equilibrium state to differ from the ensemble average \eqref{eq:ensav}. This can be efficiently computed from previous calculations simply by replacing the arbitrary operator $A$ in the ensemble average discussion by $\sum_E|E\ra\la E|A|E\ra\la E|$. When we do this, the term $\tr(A\,\Omega_E)$ remains invariant, whereas the random expectation value $\la\psi|A|\psi\ra$ becomes
\begin{equation}
 \la \psi|\left(\sum_E |E\ra\la E|A|E\ra\la E\right)|\psi\ra = \sum_E |\la E|\psi\ra|^2\,\la E|A|E \ra = \tr\left(A\,\omega_\psi\right)\,.
\end{equation}
Replacing this into \eqref{corr}, we obtain
\begin{equation}  
  \text{Prob}\left(|\tr(A\:\omega_\psi)-\tr(A\:\Omega_E)|\geq \delta a\right)\leq \frac{\tilde{\sigma}^2_A}{(\delta a)^2}\,,
\end{equation}
with the new variance $\tilde{\sigma}^2_A$ given by
\begin{equation}
\tilde\sigma^2_A\equiv\la \left(\tr(A\:\omega_\psi)-\tr(A\:\Omega_E)\right)^2\ra_\psi\,.
\end{equation}

\subsection{The use of typicality on expectation values}

Our goal is to show the above probabilities are suppressed in the dimensionality $d_E$ of the Hilbert space ${\cal H}_E$. In the process, we learn which properties the operator $A$ must satisfy to violate this conclusion. 

To study these issues we will apply {\it typicality} arguments. These follow from the phenomenon denoted as concentration of measure in the mathematical literature \cite{measure}. For our purposes, it is sufficient to point out that on a d-dimensional sphere ${\cal S}^d$ of unit radius almost all points are within geodesic distance $\frac{1}{\sqrt{d}}$ from its equator. More mathematically, given an spherical cap $C(\epsilon)$ located a distance $0< \epsilon < 1$ from the center of the sphere, the isoperimetric inequality allows us to bound its normalised measure by
\begin{equation}
  \sigma\left(C(\epsilon)\right) = \frac{\text{Area} \left(C(\epsilon)\right)}{\text{Area} \left({\cal S}^{d}\right)} \leq \left(\sqrt{1-\epsilon^2}\right)^{d+1} \leq e^{-(d+1)\epsilon^2/2}\,.
\end{equation}
It follows from this inequality that the probability for a random point on this sphere to belong to a band $B_\epsilon$ around the sphere excluding the pair of caps $C(\epsilon)$ equals
\begin{equation}
  \sigma(B_\epsilon) \geq 1 - 2\,e^{-(d+1)\epsilon^2/2}\,.
\end{equation}
Our claim follows in the limit of large $d$.

These geometric facts become relevant for the foundations of quantum statistical mechanics when we consider subspaces ${\cal H}_E$ of the Hilbert space with a large dimension. Indeed, from \eqref{eq:sphere}, a random pure state in ${\cal H}_E$ corresponds to a random point on a sphere of dimension $d=2d_E-1$ parameterised by the complex vector components $c_E$.

The application of these ideas to expectation values, i.e. functions over this sphere, is known as {\it Levy's lemma}.
\begin{lemma}{\bf Levy's lemma:}
Given a bounded function $f(\psi)$ defined over the set of pure states $\psi \in {\cal H}_E$, for any such random state $\psi$ and any $\epsilon > 0$
\begin{equation}
\text{Prob}\left(|f(\psi)-\la f(\psi)\ra_\psi| \geq \epsilon \right)\leq 2\:e^{-4c\: d_E\epsilon^2/\lambda^2}
\label{eq:levy}
\end{equation} 
where $c=(18\pi^3)^{-1}$, $d_E$ is the dimension of $\cH_E$, and $\lambda=\sup|\nabla f|$ is the Lipschitz constant of the function $f(\psi)$.
\label{levy}
\end{lemma}
See \cite{measure} for a proof.

\paragraph{Ensemble average:} As a first application of Levy's lemma, consider the random variable $X_\psi$. Its uniform probability distribution is precisely the normalised measure over the sphere describing the microstates in our microcanonical ensemble ${\cal H}_E$. Thus, expectation values of operators $A$ are functions over this sphere, i.e. $f(\psi)=\la\psi|A|\psi\ra$, satisfying 
$\la f(\psi)\ra_\psi = \text{tr}\left(A\Omega_E\right)$. The Lipschitz constant $\lambda$ controls how probable large deviations are. It was proved in \cite{popescu-1} that
\begin{equation}
  \lambda=\sup|\nabla \la\psi|A|\psi\ra| \leq 2\|A\|\,.
\label{eq:max}
\end{equation}
Thus, unless the largest eigenvalue of $A$ scales like $\sqrt{d_E}$, we can conclude that such probability is exponentially suppressed in $d_E$.

\paragraph{Time average:} The variance $\sigma_A^2(\psi)$ in \eqref{sigma} is bounded above according to \cite{reimann,short-1}
\begin{equation}\label{sigma}
\sigma_A^2(\psi)\leq \frac{\Delta_A^2}{4}\tr(\omega_\psi^2)\,.
\end{equation}
This upper bound has two pieces. The quantity $\tr(\omega_\psi^2)$ is the {\it purity} of the equilibrium state, it equals $\sum_E |c_E|^4$ and manifestly depends on the state.  The quantity $\Delta_A^2$ is the difference between the largest and the smallest eigenvalues of $A$ restricted to $\cH_E$
\begin{equation}\label{Delta_A}
\Delta_A=\max_{\psi\in\cH_E} \la \psi|A|\psi\ra-\min_{\psi\in\cH_E}\la\psi|A|\psi\ra\,.
\end{equation}
It manifestly depends on the operator $A$.

A more sophisticated application of Levy's lemma \eqref{eq:levy} in \cite{popescu-2} allows to prove  
\begin{equation}
  \label{eq:purity-b}
\text{Prob}\left(\tr(\omega_\psi^2)> \frac{4}{d_E}\right)\leq 2e^{-\tilde{c}\sqrt{d_E}}\,,
\end{equation}
where $\tilde{c}=\frac{\left(\log 2\right)^2}{72\pi^3}\approx 10^{-4}$. Thus, $\tr(\omega_\psi^2)$ is typically suppressed as $\frac{1}{d_E}$. Using this result in \eqref{chebysh_time}, we learn that the probability of having significant deviations is suppressed unless $\frac{\Delta_A}{\delta a} \sim \sqrt{d_E}$. 

We can get some intuition on this result by replacing $\delta a$ with the averaged gap $\la \delta a\ra$ between eigenvalues of the operator A. In that case, the quotient $\frac{\Delta_A}{\la \delta a\ra}\sim N(A)$ behaves likes the number of different outcomes for the operator $A$, $N(A)$. We shall return to the importance of the magnitude $N(A)$ shortly.

\paragraph{Ensemble vs time averages:} We use here the same strategy as for the ensemble average discussion, but replacing the operator $A$ by $\sum_E|E\ra\la E|A|E\ra\la E|$. Thus, we want to apply Levy's lemma \ref{levy} directly. To compute an upper bound for the corresponding Lipschitz constant, we first notice that 
\begin{equation}
  \la E |A| E\ra = \sum_a a |\la E|a\ra|^2 \quad \Rightarrow \quad |\la E |A| E\ra| \leq \sum_a |a| |\la E|a\ra|^2\,.
\end{equation}
Summing over $E$, we reach the conclusion
\begin{equation}
  |\sum_E |E\ra \la E |A| E\ra\la E| | \leq \|A \|\,.
\end{equation}
Thus, the norm of the operator $\sum_E|E\ra\la E|A|E\ra\la E|$ is bounded above by the norm of the operator $A$, $\|A\|$. This means the
probability of large deviations satisfies
\begin{equation}
  \text{Prob}\left(|\tr\left(A\,\omega_\psi\right) - \tr\left(A\,\Omega_E\right)|\geq \epsilon\right) \leq 2e^{-c\: d_E\epsilon^2/\|A\|^2}\,.
\end{equation}
As before, unless the largest eigenvalue of $A$ scales like $\sqrt{d_E}$, the probability that a random time averaged expectation value significantly differs from the ensemble average is exponentially suppressed in the dimension $d_E$ of ${\cal H}_E$. 

This last statement refers to the notion of quantum ergodicity introduced in \cite{von_Neumann} (see also the more recent discussion \cite{q-ergodic}). The latter is based on the equality between time and microcanonical averages. Here, we are saying that for almost every pure state, time average and ensemble average expectation values of operators $A$ satisfying $\|A\|^2\ll d_E$ are equal. 

To sum up,  the use of typicality allows us to argue that almost every pure state in ${\cal H}_E$ behaves like the ensemble average unless the operator we use to probe it satisfies $\|A\|^2 \sim d_E$. Similarly, the fraction of time a random pure state spends away from an equilibrium state is negligible unless $\frac{\Delta_A}{\delta a} \sim N(A) \sim \sqrt{d_E}$. Furthermore, for almost every pure state, time and microcanonical averages agree when $\|A\|^2 \ll d_E$.

\subsection{Measure of distinguishability}
\label{sec:measure}

The comparison of expectation values of an observable is not the only way to tell quantum states apart in quantum mechanics. In fact, it is easy to find examples of observables whose measurement can distinguish between different quantum states even if their expectation values are equal. Consider two spin one states  $|0\ra$ and $\frac{1}{\sqrt{2}}(|-1\ra+|1\ra)$. By construction, both states have vanishing $\sigma_z$ expectation values, however a measurement of $\sigma_z$ can easily tell the states apart. 

Notice that in this example we could distinguish both states by comparing the expectation value of $\sigma_z^2$. In quantum mechanics, given a state $\psi$ and an observable $A$, the result of its measurement is a set of eigenvalues $a$ appearing with probabilities $p_a$. It is clear that reconstructing the {\it entire probability spectrum} $\{p_a\}$ provides more information about the quantum state $\psi$ than simply measuring $\langle\psi|A|\psi\rangle$. In fact, if the observable $A$ has $N(A)$ different eigenvalues, we may need the collection of expectation values $\langle\psi| A^i|\psi\rangle$ for $i=1,\dots N(A)$ to reconstruct such probability spectrum. This reconstruction problem is called the moment problem.

This discussion motivates the notion of distinguishability introduced in the quantum information literature. The {\it distinguishability} 
of two quantum states $\rho$ and $\sigma$ using a particular observable $A$ is defined as
\begin{equation}\label{distin}
   D_A(\rho,\sigma)=\frac{1}{2}\sum_a|\tr(|a\ra\la a|\rho)-\tr(|a\ra\la a|\sigma)|\,,
\end{equation}
where $|a\ra$ are the eigenvectors of $A$.\footnote{In general, the projectors $|a\ra\la a|$ in \eqref{distin} should be replaced by a positive operator valued measure (POVM) \cite{nielsen-book}. A POVM with a finite number of outcomes is a partition of identity, $\sum_a M_a=\mathbb{I}$, where the probability of obtaining $a$ in the measurement is given by $\tr(\rho M_a)$.} Notice that this measure  is independent of the absolute value of the eigenvalues $a$ and only depends on the {\it entire} probability spectrum $\{p_a\}$. 
Thus, it is more appropriate as a measure of quantum state distinguishability than any individual expectation value. 
In particular, it is shown in \cite{nielsen-book} that the optimal probability of telling $\sigma$ and $\rho$ apart in any measurement is $\frac{1}{2}(1+D(\sigma,\rho))$. This guarantees that if $D(\rho,\sigma)$ is small, no observable can tell $\rho$ and $\sigma$ apart. We stress that the ratio $\sigma_A^2/(\langle A\rangle)^2$ could go to zero in the limit of large $d_E$ while $D_A(\rho,\Omega_E)$ may not.

This notion can be generalized to any set of observables $\mathcal{A}$,
\begin{equation}
 \label{distinset}
   D_\mathcal{A}(\rho,\sigma)=\max_{A\in\mathcal{A}}D_A(\rho,\sigma)
\end{equation}
irrespectively on whether they commute or not. In particular, if $\mathcal{A}$ includes the entire set of observables in the Hilbert space, one talks about the {\it distinguishability} of two quantum states $\rho$ and $\sigma$ as,
\begin{equation}
  D(\rho,\sigma) = \frac{1}{2}\text{tr} |\rho - \sigma|
\end{equation}
where it is understood the right hand side equals the {\it maximal} difference in probability spectra achieved over the entire set of available observables \cite{nielsen-book}.

We are interested in computing $D_A(\psi(t),\Omega_E)$, that is the distinguishability between a random pure state $\psi\in {\cal H}_E$ and the maximally mixed state $\Omega_E = \frac{\mathbb{I}_E}{d_E}$. If one identifies the conditions the observable $A$ must satisfy for $D_A(\psi(t),\Omega_E)\to 0$, one will conclude the random pure state appears entangled from the perspective of the observable $A$.
The theorem below summarizes our results. 

\begin{theorem} \label{theo}
Given a random pure state $\psi\in\cH_E$, its distinguishability $D_{\mathcal{A}}(\psi,\Omega_E)$ from the maximally mixed state $\Omega_E$ using the set of observables $\mathcal{A}$ satisfies
\begin{equation}\label{povm}
\begin{aligned}
    \text{Prob}\left( D_\mathcal{A}(\psi,\Omega_E) > \ep+\frac{N(\mathcal{A})}{2\sqrt{d_E}}\right)& \leq e^{-c \ep^2 d_E}\,, \\
\text{Prob}\left( \la D_\mathcal{A}(\psi(t),\Omega_E)\ra_t > \ep+\frac{N(\mathcal{A})}{2\sqrt{d_E}}\right)& \leq e^{-c \ep^2 d_E},
\end{aligned}
\end{equation}
for an arbitrary $\ep>0$, where $N(\mathcal{A})$ is the maximum number of outcomes of all measurements in $\mathcal{A}$.

\end{theorem}
The proof is in appendix \ref{app:b}. 

Since both statements hold for {\it any} $\epsilon > 0$, we learn that the probability for $D_\mathcal{A}(\psi,\Omega_E)$ to be larger than zero is exponentially suppressed in $d_E$ {\it unless} $N(\mathcal{A})\sim\sqrt{d_E}$. \footnote{The theorem is not useful if we consider a set ${\cal A}$ containing of the order of $\sqrt{d_E}$ different operators with individual different outcomes of order one. But it does not tell us how to implement such measurements in practice.} Furthermore, the second part of the theorem proves that typical states remain indistinguishable from equilibrium for almost all times if $N(\mathcal{A})\ll \sqrt{d_E}$. Notice that both statements hold, in particular, for an individual operator $A$. In that case, the number of different outcomes $N(A)$ refers to that single operator.

Notice theorem \ref{theo} mathematically characterises the set of operators $A$ for which a typical random pure state $\psi \in {\cal H}_E$ appears entangled. We can intuitively understand the condition on $N({\cal A})$ as saying the set of measurements in ${\mathcal A}$ is not capable of reaching the resolution required to distinguish microstates. 

When the entire set $\mathcal{A}$ has support on a finite subsystem, theorem \ref{theo} reduces to a rigorous restatement of the well-known result due to Page \cite{page}: the difference between a pure state and a mixed state in a composite Hilbert space is exponentially small unless the number of states being measured is comparable to the square root of the dimensionality of the {\it entire} Hilbert space. 
Consider a composed Hilbert space ${\cal H}_S\otimes {\cal H}_B$, where ${\cal H}_S$ and ${\cal H}_B$ are the system and bath Hilbert spaces, respectively. Let ${\cal H}_E\subseteq {\cal H}_S\otimes {\cal H}_B$ be the subspace associated with the microcanonical ensemble as defined previously. The maximally mixed state in ${\cal H}_E$ is
\begin{equation}
  \Omega_E = \frac{\mathbb{I}_E}{d_E}\,,
\end{equation}
where $\mathbb{I}_E$ is the identity matrix on ${\cal H}_E$, whereas its restriction to ${\cal H}_S$ by tracing out over ${\cal H}_B$ defines
\begin{equation}
  \Omega_S = \text{tr}_B\left(\Omega_E\right)\,.
\end{equation}
Then, from theorem \ref{theo} we obtain
\begin{equation}
   \text{Prob}\left( D_\mathcal{A}(\psi,\Omega_E) > \ep+\frac{d_S}{2\sqrt{d_E}}\right)\leq e^{-c \ep^2 d_E},
\end{equation}
since the number of outcomes is necessarily bounded by the dimension $d_S$ of the Hilbert space ${\cal H}_S$. The last statement is equivalent to the bound found in \cite{popescu-1}
\begin{equation}
\langle\|\rho_S - \Omega_S\| \rangle_S \leq \sqrt{\frac{d_S^2}{d_E}}\,.
\end{equation}

An equivalent, perhaps more explicit, way of reproducing Page's original discussion is  by considering a random quantum state for a system of $N$ spins $\sigma_i=\pm\frac{1}{2}$
\begin{equation}
  |\psi \rangle = \sum_{\{\sigma_i\}} \frac{1}{2^{N/2}}|\sigma_1,\dots ,\sigma_N\rangle\,e^{i\theta(\sigma_1,\dots,\sigma_N)}
\end{equation}
where $\theta(\sigma_1,\dots,\sigma_N)$ is a set of random phases. If only $k$ spins are measured, any observable acting on that subspace of the Hilbert space ${\cal H}_S$ will have a number of outcomes bounded above from its dimensionality $d_S=2^k$. Its expectation values can be evaluated using the density matrix $\rho_k$. Explicit calculation gives
\cite{page,nima} 
\begin{equation}
  \rho_k = \frac{1}{2^k}\sum_{\{\sigma_k\}} |\sigma_1,\dots,\sigma_k\rangle\langle\sigma_1,\dots,\sigma_k| + {\cal O}\left(2^{-N+2k}\right) =
  \Omega_k + {\cal O}\left(2^{-N+2k}\right)\,.
\end{equation}
Thus, $\rho_k$ is well approximated by the maximally entangled mixed state $\Omega_k$ unless $k\sim \frac{N}{2}$, which is equivalent to the requirement above $d_S^2\sim 2^N=d_E$.

This matches the content of Page's initial result \cite{page}, emphasizing the role played by quantum entanglement and concentration of measure \cite{measure}, which is the mathematical justification behind typicality. 

\subsection{Coarse-grained observables}
\label{sec:coarse}

Theorem \ref{theo} is relevant for first principle considerations, but it does not say whether such distinguishing measurements can be performed. For example, one could conceive the existence of operators whose spectra satisfy the requirement $N(A)\sim \sqrt{d_E}$ but have an averaged eigenvalue difference suppressed in $d_E$ itself. Since obtaining information about a system is never for free in quantum mechanics \cite{zurek}, it is an important question to discuss whether such measurements can be implemented in short time in the probe limit.

The uncertainty principle already teaches us that achieving high precision (large $N(A)$) in a measurement performed in short time requires large energies. In this section we show that any measurement performed with finite time and energy resources implements the notion of a {\it coarse-grained observable} \cite{von_Neumann}.

It is instructive to review the standard approach to measurements in field theory. Given an observable $\phi(x,t)$, its measurement is described in terms of a semiclassical process in which a classical source $J$ couples to $\phi(x,t)$ by turning on an interaction $H_{\text{int}}=\lambda\: \phi\,J \delta(x)$ at time $t$, and turning it off after an infinitesimally short amount of time. In order to keep the disturbance minimal (probe limit), one takes the limit $J\to 0$. A semi-classical measurement is not limited in precision, and correlation functions found this way are arbitrarily accurate. However, this treatment only holds approximately as a limit of a quantum process that we now describe. 

To describe a quantum measurement, we follow von Neumann \cite{VonNeumann}. We entangle our physical system with a quantum detector and consider a sharp projection on the compound state after the measurement takes place. More mathematically, assume our initial state is in a product state $|\psi\ra\otimes |\alpha\ra\in\cH_E\otimes \cH_A$, where $\cH_A$ stands for the apparatus Hilbert space. We turn on an interaction Hamiltonian $H_{\text{int}}=\lambda A\otimes J$, where $A$ is the observable we want to measure, and $J$ acts on $\cH_A$.

As before the interaction acts at a time scale $\ep$ much smaller than the inverse energy of the system such that the evolution due to the physical system Hamiltonian during the measurement can be ignored. The analogue of the probe limit is the constraint that the kick received by the ensemble due to the measurements is small, i.e. $S_{\text{int}}/\hbar=\int dt \:H_{\text{int}}/\hbar\ll E \tau/\hbar$ where $\tau=\mathcal{O}(1)$ . 

Let us expand the time evolution of the initial entangled state in the eigenbasis of $A$
 \begin{equation}
    |\Psi(t)\ra=\sum_a\psi_a(t) e^{-ia\lambda J\,t}|a\ra|\alpha\ra\,.
 \end{equation}
The measurement will distinguish the outcome $a$ from $a+\delta a$ if the apparatus wave-functions corresponding to these different eigenvalues become orthogonal at time $t$
\begin{equation}
   S(t)=\la\alpha|e^{-i \lambda (\delta a)J\,t}|\alpha\ra=0\,.
 \label{eq:measurement}
\end{equation}
This condition relates the amount of time and energy involved in a quantum measurement with resolution $\delta a$ in our observable A. It allows us to find a lower bound on the smallest resolvable eigenvalue gap $\delta a$.

\begin{theorem}\label{deltaEt}
The smallest gap $\delta a$ resolvable in a measurement of an observable $A$ in time $t$ using the interaction Hamiltonian $H_{\text{int}}=\lambda A\otimes J$  is bounded below by
\begin{equation}
\frac{\pi\hbar \la A\ra}{\la H_{int}\ra t}=\frac{\pi\hbar}{\lambda\la J\ra t}\leq \delta a.
\label{eq:gap}
\end{equation}
\end{theorem}
The proof appears in Appendix \ref{app:c}, and it is based on the existence of a universal limit on how fast a state can dynamically evolve to an orthogonal state in an isolated quantum mechanical system \cite{margolus}. 

Using \eqref{eq:gap}, we can derive the relation
\begin{equation}
N(A)\leq \frac{\Delta_A}{\delta a}\leq \frac{\Delta A}{\la A\ra}\frac{\la H_{int}\ra t}{\pi\hbar}.
 \label{eq:gap1}
\end{equation}
Measurements that distinguish microstates, i.e. $N(A)\sim \sqrt{d_E}$ always require a large kick. Thus, they always involve long times or large energy resources.\footnote{Note that for measurements that take a long time the self-Hamiltonian of the system cannot be ignored, and acts as a source of noise.}

\paragraph{Relation to coarse-grained observables :} we want to show that if the measurement action $S_{int}/\hbar$ is {\it finite}, i.e. if the amount of time and energy resources are finite, the fine grained observable $A$ effectively behaves as a coarse-grained observable due to the finite precision $\delta a$ achievable by the measurement. Indeed, given a precision $\delta a$, any observable A allows a description in terms of a coarse-grained observable
\bea
\tilde{A}=\sum_{i=0}^{N(\tilde{A})}\tilde{a}_i\Pi_i, \qquad \tilde{a}_i=\tilde{a}_0 + i\:\delta a, \qquad \Pi_i=\sum_{a\in [\tilde{a}_i, \tilde{a}_{i+1})} |a\ra\la a|.
\eea
Its eigenvalues $\tilde{a}_i$ are defined as follows : $\tilde{a}_0 = n_0\,\delta a$, where $n_0\,\delta a \leq a_{\text{min}} < (n_0+1)\,\delta a$, whereas $n_\star\,\delta a \leq a_{\text{max}} < (n_\star+1)\,\delta a$. Thus, the number of coarse-grained eigenvalues $N(\tilde{A})$ equals $n_\star-n_0+1$ and the projectors $\Pi_i$ include {\it all} the microscopic eigenvalues $a$ in the macroscopic eigen-band $(n_0+i)\delta a \leq a < (n_0+i+1)\delta a$.

Finite action $S_{\text{int}}/\hbar$ requires $\langle A\rangle \sim \delta a$. This is equivalent to $n_0, n_\star \sim \mathcal{O}(1)$. Equivalently, even if the observable $A$ has no degenerate eigenvalues, i.e. the number of microscopic outcomes is $d_E$, the number of macroscopic outcomes attainable with finite precision is order one, i.e. the observable A behaves like a coarse-grained observable $\tilde{A}$.\footnote{The connection between information loss in black holes and finite resolution of low energy measurements has been previously discussed in \cite{Balasubramanian:2006iw}.}

\section{Lessons for Quantum Gravity}
\label{sec:lessons}

In this section we discuss the implications of our results for black hole physics in an AdS/CFT context \cite{ads-cft,ads-review}. \footnote{Any precise application of our results to any quantum theory of gravity would require to extend our analysis not only to infinite dimensional Hilbert spaces but also to {\it unbounded} operators. The techniques we used here to prove equilibration using typicality have been generalized to infinite systems in \cite{Lashkari:2013iga}. Even though we believe that an understanding and mathematical formulation of this extension for QFTs and CFTs is of great importance, we shall not pursue this here.} Even though our comments are far more general, we shall use the language of ${\mathcal N}=4$ SYM with gauge group  $\SU(N)$ and its holographic dual, when appropriate. Large AdS${}_5$ black holes have both energies and entropies scaling as $N^2,\,S \sim E$. Hence, the dimension of the microcanonical ensemble $d_E$ is exponentially large in the energy of the microstates, i.e. $d_E \sim e^{aE}$, where $a$ is an order one number.

\subsection{Breaking down of effective field theory}

Our analysis finds that any quantum measurement capable of resolving the differences between random pure states and a maximally entangled one in large dimensional Hilbert spaces requires an exponentially large amount of resources, i.e. either energies or times exponential in $N^2$. Clearly, any low energy effective action does not include such degrees of freedom. For example, in AdS/CFT, the supergravity approximation deals with operators with conformal dimension $\Delta \sim{\cal O}(1)$ (no scaling in $N$), whereas classical supergravity saddle points have energies of order $N^2\sim \log d_E$. Thus, we must conclude that {\it all low energy observables} in such effective theory can only achieve this task by waiting $\log t= {\cal O}(N^2).$\footnote{Even though it is hard to prove by first principles, it is believed that any high precision measurement, as the ones we require in our discussion, lies beyond the regime of validity of an effective field theory. We thank Zohar Komargodski for emphasising this to us.} This is in agreement with the finding that 2-point correlators of light operators detect deviations from thermality at these time-scales \cite{juan}.

Equivalently, {\it random pure states appear entangled for all semiclassical gravitational probes in time polynomial in $N$}. We emphasize this statement excludes those operators built out of a large number of products of light operators. These are heavy and can distinguish microstates. Thus, these operators of the full quantum theory do {\it not} belong to the effective theory, i.e. the set of semiclassical operators ${\cal A}_\text{low}$ does {\it not} form an algebra. If one considers operators made out of the product of order $N^2$ light observables, one expects perturbation theory to break down due to the large combinatorial factors appearing in Feynman diagrams, similar to the appearance of non-planar effects in non-abelian gauge theories \cite{oldwitten}.\footnote{Perturbation theory in gravity is an expansion in Newton's constant $G_N$. When computing correlators in black hole microstates, the perturbative expansion is expected to be in $1/S$, where $S=A/(4G_N)$ is the standard Hawking-Bekenstein formula. For correlators involving $m\sim N^2 \sim S$ insertions, one expects such perturbative arguments to break down.}

Both, our quantum mechanical results in section \ref{sec:3} and the arguments above suggest that the number of operator insertions, $N$ in equation \eqref{eq:effect} and their conformal dimensions $\Delta_i = \Delta\left({\cal O}_i\right)$ play a similar role to the number of outcomes in our quantum mechanics discussion \cite{essay,babel,nima,review,papadodimas}. We postpone a precise mathematical formulation of this problem to future work. But we stress this expectation holds in effective field theory. This is because there exists a map between low energy operators and particle excitations in this regime. Thus, the number of operator insertions in a given correlator describes how large the dimension of the perturbative Fock space is being explored. In this respect, the argument is analogous to our spin discussion in section \ref{sec:measure}.

\subsection{Effective emergence of locality}

Holography allows us to apply our quantum mechanical results to gravitational physics. In particular, we know the low energy sector of the bulk physics involves classical gravity, where {\it locality} is manifest. Our work proves that any operator capable of distinguishing a random pure state from a mixed state does not belong to the low energy effective action. Thus, inverting the logic presented in the last subsection, we can say that the notion of bulk locality emerges when we restrict the entire set of observables to ${\cal A}_\text{low}$.

We finish this work emphasizing that the restriction to the subspace of observables compatible with an effective notion of bulk locality is analogous to the recent proposal by Papadodimas \& Raju  \cite{papadodimas} reconstructing {\it local bulk operators} in the interior of a black hole from boundary quantum data, building on the seminal work in \cite{kabat}.

\paragraph{Emergence of bulk locality and relation to Papadodimas \& Raju : } According to \cite{kabat}, local bulk operators $\phi_{\text{CFT}}(t,\Omega,z)$ in the exterior of a black hole can be constructed as
\begin{equation}
  \phi^i_{\text{CFT}}(t,\Omega,z) = \sum_{\vec{\ell},\omega>0} \left({\cal O}^i_{\vec{\ell},\omega}f_{\vec{\ell},\omega}(t,\Omega,z) + \text{h.c.}\right) 
 \label{eq:bulkop}
\end{equation}
where $(t,\Omega,z)$ are boundary labels which can be interpreted as bulk AdS coordinates, ${\cal O}^i_{\vec{\ell},\omega}$ are the Fourier modes of a boundary local operator on the sphere and $f_{\vec{\ell},\omega}(t,\Omega,z)$ are appropriately chosen functions. Such bulk operators can be constructed, order by order in a $\frac{1}{N}$ expansion \cite{kabat-exp}.

Recently, this construction was claimed to be extended for operators probing the {\it interior} of a black hole \cite{papadodimas}. In this case, the field \eqref{eq:bulkop} is replaced by the mode expansion
\begin{equation}
  \phi^i_{\text{CFT}}(t,\Omega,z) = \sum_{\vec{\ell},\omega>0} \left({\cal O}^i_{\vec{\ell},\omega}g^{(1)}_{\vec{\ell},\omega}(t,\Omega,z) + \tilde{{\cal O}}^i_{\vec{\ell},\omega}g^{(2)}_{\vec{\ell},\omega}(t,\Omega,z)\right) 
 \label{eq:bulkop-1}
\end{equation}
where $g^{(a)}_{\vec{\ell},\omega}(t,\Omega,z)$ $a=1,2$ can be found in \cite{papadodimas-0}. The expansion \eqref{eq:bulkop-1} involves two sets of operator modes : ${\cal O}^i_{\vec{\ell},\omega}$ as before and the {\it mirror operators} $\tilde{{\cal O}}^i_{\vec{\ell},\omega}$. The latter were defined in \cite{papadodimas} as those satisfying
\begin{equation}
  \tilde{{\cal O}}^i_{\vec{\ell},\omega}\,A_\alpha |\psi \rangle = A_\alpha\,e^{-\beta\omega}\,{\cal O}^i_{-\vec{l},-\omega} |\psi \rangle
 \label{eq:mirror}
\end{equation}
where $A_\alpha$ belongs to the subset of local boundary observables 
\begin{equation}
  A_\alpha = \sum_P \alpha(P)\left({\cal O}^i_{\vec{\ell},\omega}\right)^{P(i,\omega,\vec{\ell})}
\end{equation}
belonging to the low energy sector ${\cal A}_{\text{low}}$. This last condition was more precisely stated in \cite{papadodimas} by requiring the functions $P(i,\omega,\vec{\ell})$ to satisfy
\begin{equation}
  \sum_{i,\vec{\ell},\omega} P(i,\omega,\vec{\ell})\,\omega \ll N
\label{eq:energy}
\end{equation}
Thus, by construction, $A_\alpha \in {\cal A}_\text{low}$ in our previous discussion. Papadodimas \& Raju showed mirror operators exist and that they depend on the specific state $|\psi\rangle$ where they act on \cite{papadodimas}.

The defining equation \eqref{eq:mirror} guarantees that boundary correlators are compatible with thermal behaviour. Furthermore, all operators $A_\alpha\in {\cal A}_\text{low}$ satisfy $A_\alpha |\psi \rangle \neq 0$. Thus, states $|\psi\rangle$ do appear entangled for the subset of observables $A_\alpha$ satisfying \eqref{eq:energy}. 

The connection with our work is as follows.  We characterised the set of operators that do not distinguish random pure states from the maximally entangled state $\Omega_E$ in a microcanonical ensemble ${\cal H}_E$. Our analysis was in the entire quantum theory and our only assumption on the spectrum was the existence of non-degenerate gaps to exclude integrable systems. \footnote{The non-degenerate gap assumption can be further relaxed \cite{short-2}.} When we embed our discussion into a holographic theory, the latter is believed to have an spectrum of relatively sparsed ${\cal O}(1)$ excitations separated by a large gap from a densed set of heavy states (black hole microstates) \cite{joe}. In these theories, it is natural to identify the subset of operators ${\cal A}_\text{low}$ with those  satisfying the more precise constraint \eqref{eq:energy}. These satisfy
\begin{equation}
 D_{A_\alpha}(\psi, \Omega_E) \to 0 \quad \quad  \forall\,A_\alpha \in {\cal A}_\text{low} 
\label{eq:relation}
\end{equation}
Thus all operators considered in \cite{papadodimas} are expected to satisfy \eqref{eq:relation}.

These statements can be made more explicit for a system of N spins. If we restrict ourselves to the subset of observables acting on subsystems of $k$ spins, where $k\ll \frac{N}{2}$, we already showed in section \ref{sec:3} that random pure states appear entangled. Thus, in this example it is clear that the set of operators $A_\alpha$ to be considered are those satisfying the property \eqref{eq:relation}.

A further outcome of this discussion is that the construction in \cite{papadodimas} may be less state dependent than what it appears, since all operators $A_\alpha$ see typical random pure states $|\psi \rangle \in{\cal H}_E$ as $\Omega_E$. The latter follows from typicality and as such, it is necessarily a probabilistic statement at this stage that requires further investigation.


Having discussed the connection between our results and the emergence of locality, we comment below on the implications for complementarity and the existence of a black hole interior. Some of the considerations below are necessarily similar to the ones also discussed in \cite{papadodimas}.

\paragraph{Complementarity :}  One of the main conceptual issues in black hole physics regards the compatibility of black hole complementarity \cite{complementarity} with the preservation of bulk locality. One consequence of the former is that the degrees of freedom inside and outside of the black hole are {\it not} independent. The possible tension between both concepts can easily be phrased. Low energy bulk observers using an EFT description would expect correlation functions between bulk operators in the interior $(z_{\text{int}})$ and exterior $(z_{\text{ext}})$ of the black hole to vanish
\begin{equation}
  [\phi_{\text{EFT}}(t,\Omega,z_{\text{int}}),\,\phi^\prime_{\text{EFT}}(t^\prime,\Omega^\prime,z_{\text{ext}})]_{\text{EFT}} = 0\,.
\label{eq:comm}
\end{equation}
This is because in the EFT approximation, one is doing a quantum field theory calculation in a black hole background and the points $z_{\text{int}}$ and $z_{\text{ext}}$ are causally disconnected. From the exact quantum theory of gravity perspective, one is probing the operator equation \eqref{eq:comm} in its heavy sector using the CFT operators \eqref{eq:bulkop-1}. That is, one is considering correlators of the form
\begin{equation}
  \Phi_i (z_{\text{int}},z_{\text{ext}}) \equiv \langle \psi_i |  [\phi_{\text{CFT}}(t,\Omega,z_{\text{int}}),\,\phi^\prime_{\text{CFT}}(t^\prime,\Omega^\prime,z_{\text{ext}})] |\psi_i \rangle 
\label{eq:comm-exact}
\end{equation}
with $\psi_i \in {\cal H}_E$ being the black hole microstates. One expects that such correlators are well approximated by the thermal correlator. Consequently,
\begin{equation}
\begin{aligned}
  \Phi_i (z_{\text{int}},z_{\text{ext}}) & \approx \langle \psi_{\text{BH}} |  [\phi_{\text{CFT}}(t,\Omega,z_{\text{int}}),\,\phi^\prime_{\text{CFT}}(t^\prime,\Omega^\prime,z_{\text{ext}})] |\psi_{\text{BH}} \rangle \\
 & \approx [\phi_{\text{EFT}}(t,\Omega,z_{\text{int}}),\,\phi^\prime_{\text{EFT}}(t^\prime,\Omega^\prime,z_{\text{ext}})]_{\text{EFT}} = 0\,.
\end{aligned}
\end{equation} 
This would seem to violate complementarity given the expected non-independence of interior and exterior black hole degrees of freedom.

Our results in sections \ref{sec:measure} and \ref{sec:coarse} provide mathematical evidence for the existence of a conceptual framework to understand in which sense both statements can be compatible. By first principles, the commutator \eqref{eq:comm-exact} is {\it non-vanishing}. For the subset of operators in ${\cal A}_{\text{low}}$ for which an effective local bulk description exists, deviations from thermal behaviour are exponentially suppressed in the entropy of the black hole
\begin{equation}
  \Phi_i (z_{\text{int}},z_{\text{ext}}) = {\cal O}\left(e^{-S_\text{BH}}\right) \quad \quad \forall \phi_{\text{CFT}},\,\phi^\prime_{\text{CFT}} \in {\cal A}_\text{low}
\end{equation}
Thus, even though these corrections are difficult to measure, particularly so in any semiclassical approximation, they are still responsible for the non-vanishing of the exact quantum gravity answer. For the subset of operators for which no notion of locality exists, there is no reason to expect the commutator to vanish. Thus, there was no tension to begin with.

In quantum mechanics, it is possible to construct {\it commuting coarse-grained} operators from microscopic non-commuting ones, starting with the momentum and position operators \cite{von_Neumann}, explaining why classical apparatus can measure them simultaneously. What we are saying here is that not only low energy gravity probes appear to be entangled, but they may also appear to commute due to their coarse-grained nature, even though this is not true for the exact microscopic operators.

\paragraph{Existence of a geometric black hole interior :}  There are two main reasons why we associate black holes with thermal states in the AdS/CFT correspondence. First, because in its euclidean path integral formulation, the euclidean AdS black hole is the dominant saddle contribution (at large enough temperatures) and matches the CFT expectation values of a thermal density matrix \cite{witten}. Second, because eternal AdS black holes are believed to be described by the entangled state \cite{juan}
\begin{equation}
  \rho =  |\psi_{\text{BH}}\rangle \langle \psi_{\text{BH}}|\,, \quad \text{with} \quad
  |\psi_{\text{BH}}\rangle = \sum_{n\in{\cal H}} \frac{e^{-\beta E_n/2}}{\sqrt{Z}} |n\rangle_{{\cal H}_L} \otimes  |n\rangle_{{\cal H}_R} 
\end{equation}
where ${\cal H}_L$ and ${\cal H}_R$ stand for the two isomorphic Hilbert spaces defined on each of their two conformal boundaries.
An asymptotic observer living on ${\cal H}_R$ will only measure observables ${\cal O}_R$ acting on ${\cal H}_R$. By construction, such expectations values are captured by the reduced density matrix obtained by tracing over ${\cal H}_L$
\begin{equation}
  \rho_R = \text{tr}_{{\cal H}_L} \rho = \frac{1}{Z}\sum_{n\in{\cal H}_R} e^{-\beta E_n} |n\rangle\langle n|\,.
\end{equation}
Thus, these expectation values are manifestly thermal
\begin{equation}
  \text{tr}_{{\cal H}_L\otimes {\cal H}_R} \left(\rho\,{\cal O}_R\right) = \text{tr}_{{\cal H}_R}\left(\rho_R\,{\cal O}_R\right) = \frac{1}{Z}\sum_{n\in {\cal H}_R} e^{-\beta E_n} \langle n |{\cal O}_R| n \rangle\,.
\end{equation}

One natural question this eternal AdS black hole description raises is : what is the holographic dual of the quantum state $\rho_R$ ? The reason to ask this question is because there are many different quantum states in ${\cal H}_L\otimes {\cal H}_R$ giving rise to the same density matrix $\rho_R$ capturing the physics of a single asymptotic observer. This suggests that $\rho_R$ only describes the exterior of the AdS event horizon \cite{Czech:2012bh,Hubeny:2012wa,Czech:2012be,mark}. 

In fact, this is also heuristically suggested by euclidean path integral considerations. The euclidean black hole is an smooth geometry constructed from the lorentzian geometry by cutting the horizon, {\it removing its interior} and gluing it in an smooth way. It is indeed true that such euclidean saddle knows about the full entropy of the system, but it does not know anything about the specific microstates of the system \cite{Simon:2009mf}\footnote{For a recent discussion on the relation of microstates to thermodynamics and euclidean path integral consideraions, see \cite{recent-samir}.}. This perspective suggests that information about the black hole microstates is "hidden" in the interior, but does such interior allow for a local bulk description ?
 
The arguments used to reconcile complementarity with EFT locality expectations already suggest the answer depends on the observables one studies. Whenever the effective black hole geometry is reliable, that is when we probe it with operators $\phi_{\text{CFT}}\in {\cal A}_\text{low}$, we can obviously not distinguish between the black hole and the microstates. There should exist some effective notion of interior. This is made more precise in \cite{papadodimas}. If an asymptotic observer attempts to improve on this situation, he/she must either wait for a long time or consider correlators involving operators {\it not} in ${\cal A}_\text{low}$. These operators can carry very large energies, could distinguish among black hole microstates and will generically have large variances in the ensemble of microstates. Thus, there is no well defined notion of locality for these.

\section*{Acknowledgements}
We would like to thank Vijay Balasubramanian, Jos\'e Barb\'on, Micha Berkooz, Ignacio Cirac, Patrick Hayden, Seth Lloyd, Simon Ross, James Sully, and Leonard Susskind for discussions. JS would like to thank University of Oviedo and the organizers and participants of the  {\it Quantum Information} workshop in Benasque, {\it Black holes in String Theory} workshop in Ann Arbor and the  COST-WIS wokshop on {\it Black holes and quantum information} in the Weizmann Institute for hospitality during different stages of this work. The work of NL is supported in part by the Natural Sciences and Engineering Research Council of Canada. The work of JS was partially supported by the Engineering and Physical Sciences Research Council (EPSRC) [grant number EP/G007985/1], the Science and Technology Facilities Council (STFC) [grant number ST/J000329/1] and a {\it Visiting Professor} fellowship at the Universidad de Oviedo.

\appendix

\section{Ensembles, averages \& variances}
\label{app:a}

We briefly review the definition of the ensemble and time averages discussed in section \ref{sec:3.1}.

\paragraph{Ensemble averages :} The most general parameterisation of the ensemble of states $\psi\in{\cal H}_E$ is in terms of the arbitrary set of coefficients $c_E$ describing the expansion of {\it any} state $\psi$ in the eigen-energy basis.
\begin{equation}
  |\psi(0)\rangle = \sum_E c^{\,\psi}_E |E\rangle \quad \text{with} \quad \sum_E |c^{\,\psi}_E|^2 = 1
\label{eq:gen-sta}
\end{equation}
Given an observable ${\cal O}$, its expectation value on such a random state equals.
\begin{equation}
  \langle \psi(0)|{\cal O}|\psi(0)\rangle = \sum_{E,E^\prime} c^{\,\psi}_E\,c^{\psi\,\star}_{E^\prime}\langle E^\prime|{\cal O}|E\rangle
\end{equation}
The averaged expectation value $\langle {\cal O} \rangle_E$ is defined as,
\begin{equation}
 \langle {\cal O} \rangle_E = \int d\vec{c}^{\, \psi} \,
     \sum_{E\,, E^\prime} c_E^\psi c_{E^\prime}^{\psi\, *}
       \langle E^\prime| {\cal O} |E\rangle
\label{eq:av1}
\end{equation}
where the measure is normalised to one, i.e. $\int d\vec{c}^{\, \psi}=1$.

This measure satisfies the property \cite{mukund}
\begin{equation}
  \int d\vec{c}^{\, \psi} c_E^\psi c_{E^\prime}^{\psi\, *} = \frac{1}{d_E}\delta_{EE^\prime}\,.
\label{eq:id1}
\end{equation}
Equivalently, the average dephases the state. Using \eqref{eq:id1} in \eqref{eq:av1}, 
\begin{equation}
 \la {\cal O} \ra_E = \frac{1}{d_E}\text{tr} \,{\cal O} = \text{tr}\left({\cal O}\,\Omega_E\right) \quad \text{with} \quad \Omega_E = \frac{\mathbb{I}_E}{d_E}
\end{equation}
reproducing the statement \eqref{eq:ensav} in the main text. 

We define the variance $\sigma^2_{{\cal O}}$ as for any other ensemble.
\begin{equation}
  \sigma^2_{{\cal O}}\equiv \left\langle \left(\langle\psi |{\cal O}|\psi\rangle -  \langle {\cal O}\rangle_E\right)^2 \right\rangle_E
\end{equation}
This satisfies $\sigma^2_{{\cal O}} = \la\la\psi|{\cal O}|\psi\ra^2\ra_E - \frac{\left(\text{tr} \,{\cal O}\right)^2}{d_E^2}$ which equals
\begin{equation}
  \sigma^2_{{\cal O}} = \int\, d\vec{c}^{\, \psi} \left( \sum_{E_1\,, E_2\,, E_3\,, E_4}
     c_{E_1}^\psi c_{E_2}^{\psi\, *} c_{E_3}^\psi c_{E_4}^{\psi\, *} \,
       \langle E_2 | {\cal O} | E_1 \rangle \, \langle E_3 | {\cal O} | E_4 \rangle\right) -  \frac{\left(\text{tr} \,{\cal O}\right)^2}{d_E^2}
\end{equation}  
Using the integral identity \cite{mukund}
\begin{equation}
 \int\, d\vec{c}^{\,\psi} |c_E^\psi|^2 \, |c_{E^\prime}^\psi|^2 = \frac{1 + \delta_{EE^\prime}}{d_E(d_E+1)}\,,
\end{equation}
in the first term and taking into account that the same dephasing as in \eqref{eq:id1} occurs, we conclude
\begin{equation}
  \sigma^2_{{\cal O}} = \frac{1}{d_E+1}\left(\frac{\tr({\cal O}^2)}{d_E}-\frac{\tr({\cal O})^2}{d_E^2}\right)\,.
\label{eq:var1}
\end{equation}

There exists a different commonly used possibility to describe ensemble averages using unitary integrals. In this case, one averages the expectation values of a given operator ${\cal O}$ over the entire set of states obtained from $\psi$ by a unitary transformation $U$. The averaged expectation value equals
\begin{equation}
  \langle {\cal O}\rangle_\psi = \int dU \langle\psi\otimes U|{\cal O} | U\otimes\psi \rangle =  \frac{1}{d_E}\text{tr} \,{\cal O}
\label{eq:aveU}
\end{equation}
where we used the identity \cite{Collins}
\begin{equation}
  \int dU \,U_{ij}\,U^\star_{kl} = \frac{1}{d_E} \delta_{ik}\delta_{jl}\,.
\end{equation}
Thus, both ensemble averages are equal and match the microcanonical average by construction.

The variance is defined as before, but using the unitary integral average
\begin{equation}
  \sigma_{{\cal O}}^2\equiv\la \left(\la\psi|{\cal O}|\psi\ra-\tr({\cal O}\,\Omega_E)\right)^2\ra_\psi\,.
\label{eq:varU}
\end{equation}
The only non-trivial calculation left is
\begin{equation}
\begin{aligned}
\la \la \psi |{\cal O}|\psi\ra^2\ra_\psi&=\int dU\:\la \psi\otimes\psi|U\otimes U\:({\cal O}\otimes {\cal O})\:U^\dagger\otimes U^\dagger|\psi\otimes\psi\ra \\
&=\frac{1}{d_E^2-1}\la\psi\otimes\psi|\left((\tr {\cal O})^2-\frac{1}{d_E}\tr({\cal O}^2)\right)\mathbb{I} \\
&+\left(\tr({\cal O}^2)-\frac{1}{d_E}(\tr {\cal O})^2\right)\mathbb{S}|\psi\otimes\psi\ra \\
&=\frac{1}{d_E(d_E+1)}\left[\tr({\cal O}^2)+(\tr {\cal O})^2\right]
\end{aligned}
\end{equation}
where we used the identity \cite{Collins}
\begin{multline}
  \int dU\,U_{i_1j_1}\,U_{i_2j_2}\,U^\star_{i^\prime_1j^\prime_1}\,U^\star_{i^\prime_2j^\prime_2} = \frac{1}{d_E^2-1}\left(\delta_{i_1i^\prime_1}\delta_{i_2i^\prime_2}\delta_{j_1j^\prime_1}\delta_{j_2j^\prime_2} + \delta_{i_1i^\prime_2}\delta_{i_2i^\prime_1}\delta_{j_1j^\prime_2}\delta_{j_2j^\prime_1} \right) \\
  - \frac{1}{d_E\left(d_E^2-1\right)}\left(\delta_{i_1i^\prime_1}\delta_{i_2i^\prime_2}\delta_{j_1j^\prime_2}\delta_{j_2j^\prime_1} + \delta_{i_1i^\prime_2}\delta_{i_2i^\prime_1}\delta_{j_1j^\prime_1}\delta_{j_2j^\prime_2} \right)
\end{multline}
and the definition of the swap operator $\mathbb{S}$ as $\mathbb{S}|i,j\ra=|j,i\ra$. Plugging this back into \eqref{eq:varU}, we reproduce the variance \eqref{eq:var1} computed above.

\paragraph{Time averages :} In this ensemble, we consider the time evolution of the most general pure state $\psi\in {\cal H}_E$ in \eqref{eq:gen-sta}
\begin{equation}
|\psi(t)\ra=\sum_E\:c^\psi_E\:e^{-i E t}|E\ra\,,
\label{eq:timeevol}
\end{equation}
viewing any expectation value $\la\psi(t)|{\cal O}|\psi(t)\ra$ as a random variable uniformly distributed over $t\in [0,\,\infty)$, so that the time averaged expectation value is defined by
\begin{equation}
  \la {\cal O}\ra_t = \lim_{T\to\infty}\frac{1}{T} \int_0^Tdt\:\la \psi(t)|{\cal O}|\psi(t)\ra = \sum_E |c_E^\psi|^2 \la E|{\cal O}|E\ra = \tr\left({\cal O}\,\omega_\psi\right)\,,
\end{equation}
where
\begin{equation}
  \omega_\psi\equiv \langle \psi(t)\rangle_t = \lim_{T\to\infty}\frac{1}{T} \int_0^Tdt\:|\psi(t)\ra\la\psi(t)| = \sum_E |c_E|^2 |E\rangle\langle E|
\end{equation}
is the time averaged of the density matrix $\rho^\psi(t) = |\psi(t)\ra\la\psi(t)|$ constructed out of the initial pure state \eqref{eq:timeevol}.

The variance is defined as in any other ensemble as
\begin{equation}
  \sigma_{{\cal O}}^2(\psi)=\la \left(\tr({\cal O}\:\psi(t))-\la {\cal O}\ra_t \right)^2\ra_t\,.
\end{equation}
Since $\la {\cal O}\ra_t$ is diagonal, this variance only has contributions from the off-diagonal matrix elements of the operator ${\cal O}$ in the energy eigenbasis, so that
\begin{equation}
\begin{aligned}
  \sigma_{{\cal O}}^2(\psi) &=\left\la \left(\sum_{E_1,E_2}c_{E_1}c_{E_2}^*\:e^{i(E_2-E_1)t}\la E_2|{\cal O}|E_1\ra\right)^2\right\ra_t\ \\
  &=\sum_{E_1\neq E_2,E_3\neq E_4}c_{E_1}c_{E_2}^*c_{E_3}c_{E_4}^*\:\la e^{i(E_2-E_1+E_4-E_3)t}\ra_t\la E_2|A|E_1\ra\la E_4|A|E_3\ra\,.
\end{aligned}
\end{equation}
The time average forces the exponent $E_2-E_1+E_4-E_3$ to vanish. It is here where the assumption on the Hamiltonian having non-degenerate energy gaps enters. Assuming the latter and using that $E_1\neq E_2$ and $E_3\neq E_4$, we conclude the only terms surviving the intrinsic dephasing due to the time average are $E_1=E_4$ and $E_2=E_3$, so that the variance equals
\begin{equation}
   \sigma_{{\cal O}}^2(\psi) = \sum_{E,E'}|c_E|^2|c_{E'}|^2\:|\la E'|A|E\ra|^2-\sum_E |c_E|^4\:|\la E|A|E\ra|^2\,,
\end{equation}
where we added and subtracted the last term to write the final more symmetric expression.

\section{Proof of theorem \ref{theo}}
\label{app:b}

As most statements regarding probabilities in Hilbert spaces of large dimensionality, the strategy of the proof relies in the use of Levy's lemma \ref{levy}. First, choose the function $f(\psi)$ in Levy's lemma to be the distinguishability of $\psi\in {\cal H}_E$ from the maximally mixed state $\Omega_E=\mathbb{I_E}/d_E$ using the set of observables $\mathcal{A}$, $D_{\cal A}(\psi,\Omega_E)$.

We need to compute the ensemble average $\langle D_\mathcal{A}(\psi,\Omega_E)\rangle_\psi$. Due to \eqref{distinset}, we can bound this average by.
\begin{equation}
  \la D_\mathcal{A}(\psi,\Omega_E)\ra_\psi\leq \sum_{A\in \mathcal{A}}\la D_A(\psi,\Omega_E)\ra_\psi
\label{eq:easyb}
\end{equation}
This allows us to restrict the application of Levy's lemma to a single observable $A\in {\cal A}$. Thus, we need to compute $\langle D_A(\psi,\Omega_E)\rangle_\psi$ and its associated Lipschitz constant $\lambda$. 

Since $\| D_A(\psi,\Omega_E) \| \leq 1$, because each individual state probability is bounded by unity, we conclude using \eqref{eq:max}  that $\lambda \leq 2$. Next, we find an upper bound on $\langle D_A(\psi,\Omega_E)\rangle_\psi$ where the observable $A\in \mathcal{A}$ is described by POVM elements $M_a$.
\begin{equation}
\label{da}
\begin{aligned}
\la D_A(\psi,\Omega_E)\ra_\psi&=\frac{1}{2}\sum_a\la|\tr(M_a\psi)-\tr(M_a\Omega_E)|\ra_\psi \\
&= \frac{1}{2}\sum_a\la\sqrt{\tr(M_a^E(\psi-\Omega_E))^2}\ra_\psi \\
&\leq \frac{1}{2}\sum_a\sqrt{\la\tr(M_a^E(\psi-\Omega_E))^2\ra_\psi} \\
&\leq \frac{1}{2}\sum_a\sqrt{\la\tr(M_a^E\psi)^2\ra_\psi-\tr(M_a^E\Omega_E)^2},
\end{aligned}
\end{equation}
$M^E_a=\Pi^E M_a\Pi^E$ is the restriction of $M_a$ to $\cH_E$, we used the Cauchy-Schwarz inequality in the third line and used \eqref{eq:aveU} to write $\la \psi\ra_\psi=\Omega_E$. \footnote{If none of the observables in $\mathcal{A}$ take states outside of $\cH_E$ there is no need to introduce $M_a^E$.} Using \eqref{eq:var1},
\begin{equation}
\begin{aligned}
\la D_A(\psi,\Omega_E)\ra_\psi&\leq \frac{1}{2}\sum_a\sqrt{\frac{\tr((M_a^E)^2)}{d_E(d_E+1)}} \\
&\leq \frac{N(A)}{2\sqrt{d_E+1}}\leq \frac{N(A)}{2\sqrt{d_E}}
\end{aligned}
\end{equation}
where we used $\tr((M_a^E)^2)\leq \|M_a\|^2\tr(\Pi^E)\leq d_E$, and $N(A)$ is the total number of outcomes, i.e. range of the sum over $a$. Inserting this bound in \eqref{eq:easyb}, we find
\begin{equation}
\la D_\mathcal{A}(\psi,\Omega_E)\ra_\psi\leq \sum_{A\in \mathcal{A}}\la D_A(\psi,\Omega_E)\ra_\psi\leq \frac{N(\mathcal{A})}{2\sqrt{d_E}}
\end{equation}
where $N({\cal A})$ is the total number of different outcomes in the full set ${\mathcal A}$.

Inserting this bound in Levy's lemma allows to prove the first of the probability statements in theorem \ref{theo}.

The second part of theorem \ref{theo} is proved by repeating the argument above for the time-averaged distinguishability as the function of $\psi$ in Levy's lemma, and replacing $M_a$ by $e^{iH t}M_a e^{-iH t}$:
\bea
\la \la D_\mathcal{A}(\psi(t),\Omega)\ra_t\ra_\psi=\lim_{T\to\infty}\frac{1}{T}\int dt\:\la D_\mathcal{A}(\psi(t),\Omega)\ra_\psi\leq \frac{N(\mathcal{A})}{2\sqrt{d_E}}.
\eea
This finishes the proof of of indistinguishability from equilibrium for almost all states at almost all times.

\section{Proof of theorem \ref{deltaEt}}
\label{app:c}

Let us expand the initial state of the apparatus $|\alpha\rangle = \sum_j \alpha_j |j\rangle$ in the eigenbasis $|j\rangle\in \cH_A$ of the operator $J$ appearing in the interaction Hamiltonian $H_{\text{int}}=\lambda A\otimes J$. The proof first consists in establishing a bound involving the real and imaginary components of $S(t)$
\begin{equation}
   S(t)=\la\alpha|e^{-i \lambda (\delta a)J\,t}|\alpha\ra\,.
\end{equation}
Direct calculation allows us to find
\begin{equation}
\begin{aligned}
    \text{Re}\,S(t) &= \sum_j |\alpha_j|^2\cos((\delta a)\lambda j t) \\
    &\geq \sum_j|\alpha_j|^2\left(1-\frac{2}{\pi}\left((\delta a)\lambda j t+\sin(\lambda(\delta a)j t)\right)\right) \\
    &=1-\frac{2(\delta a)\lambda \la J\ra t}{\pi}+\frac{1}{\pi}\:\text{Im}\,S(t),
\end{aligned}
\end{equation}
where we used $\cos(x)\geq 1-\frac{2}{\pi}(x+\sin(x))$ in the second line. When we impose the orthonormality condition $S(t)=0$ \eqref{eq:measurement},  to guarantee the distinction between two different eigenvalues of the operator $A$ separated by $\delta a$, we derive the inequality
\begin{equation}
 \label{Margo}
    \frac{\pi}{\lambda\la J\ra t}\leq \delta a\,, 
\end{equation}
proving \eqref{eq:gap} holds.

\end{document}